# A focal plane detector design for a wide-band Laue-lens telescope

Ezio Caroli[1], Natalia Auricchio[1,2], Lorenzo Amati[1], Yuriy Bezsmolnyy[3], Carl Budtz-Jørgensen[4], Rui M. Curado da Silva[5], Filippo Frontera[2,1], Alessandro Pisa[2], Stefano Del Sordo[6], John B. Stephen[1], Giulio Ventura[1]

*(1) INAF/IASF-Bologna, Italy*
*(2) Dipartimento di Fisica, Università di Ferrara, Italy*
*(3) Semiconductor Materials and Instruments Laboratory Ltd , Ukraine*
*(4) Danish National Space Centre, Copenhagen, Denmark*
*(5) Departamento de Fisica, Universidade de Coimbra, Portugal*
*(6) INAF/IASF-Palermo, Italy*

**Abstract.** The energy range above 60 keV is important for the study of many open problems in high energy astrophysics such as the role of Inverse Compton with respect to synchrotron or thermal processes in GRBs, non thermal mechanisms in SNR, the study of the high energy cut-offs in AGN spectra, and the detection of nuclear and annihilation lines. Recently the development of high energy Laue lenses with broad energy bandpasses from 60 to 600 keV have been proposed for a Hard X ray focusing Telescope (HAXTEL) in order to study the X-ray continuum of celestial sources. The required focal plane detector should have high detection efficiency over the entire operative range, a spatial resolution of about 1 mm, an energy resolution of a few keV at 500 keV and a sensitivity to linear polarization. We describe a possible configuration of the focal plane detector based on several CdTe/CZT pixelated layers stacked together to achieve the required detection efficiency at high energy. Each layer can operate both as a separate position sensitive detector and polarimeter or work with other layers to increase the overall photopeak efficiency. Each layer has a hexagonal shape in order to minimize the detector surface required to cover the lens field of view. The pixels would have the same geometry so as to provide the best coupling with the lens point spread function and to increase the symmetry for polarimetric studies.

## 1. Laue lenses for high energy X-Rays

The energy band > 60 keV is of key importance in the study the radiation emission mechanisms of the celestial sources, their nature, their evolution with time and their contribution to the Cosmic X-ray Background (CXB). The development of focusing optics at these energies is necessary to increase the statistical quality of the high energy spectra of weak celestial sources. Recently high energy Laue lenses[Von Ballmoos P. and Smither R. K., 1994; Pisa et al., 2005] with broad energy band-passes from 60 to 600 keV have been proposed for use in a Hard X ray focusing Telescope (HAXTEL): a multi–lens configuration appears the most efficient and compact solution. To extend the telescope band-pass to lower energies, multilayer optics and/or coded mask telescopes, depending on the science goals, can complement the lenses.

The nominal configuration of the multi-lens telescope [Frontera et al., 2005] includes a Low Energy Lens (LEL) with a passband from 60 to 200 keV, containing 4 High Energy Lenses (HEL) with an operational energy range from 150 to 600 keV, in the hole left free from the LEL. The assumed focal length is 50 m and the expected Point Spread function (PSF) radius on axis is ≈ (2.63±0.03) cm for LEL and ≈ (2.33±0.03) cm for HEL at 90% collected photons (Fig. 1).

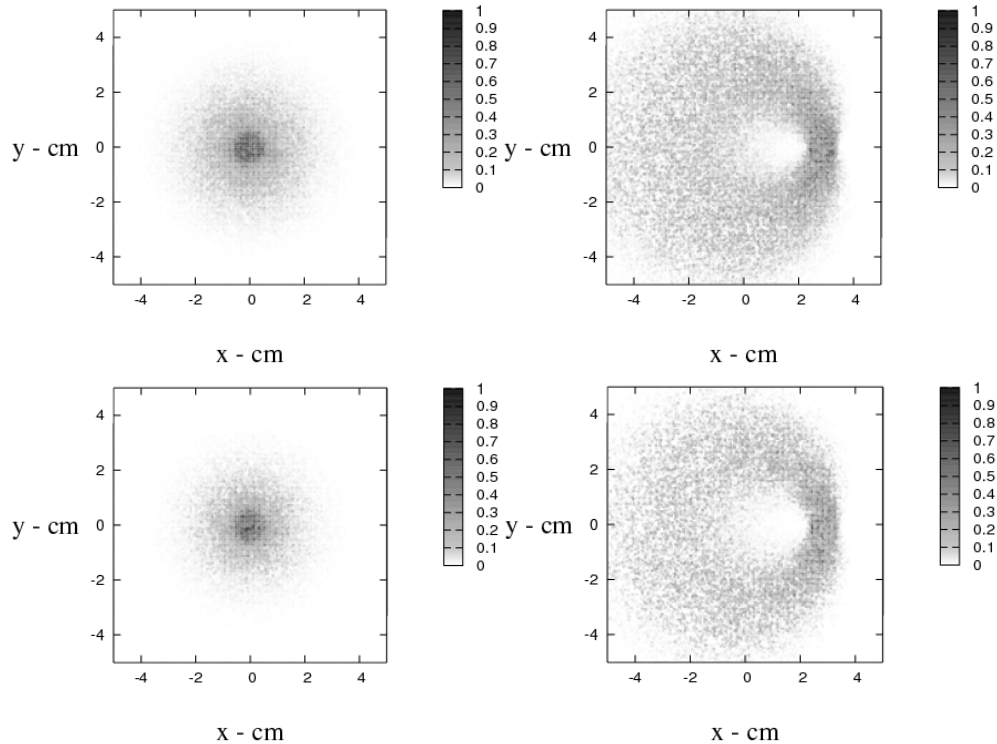

Fig. 1. The Point Spread function (PSF) on axis (left) and 2' off axis (right) at 120 keV for the LEL (top) and the HEL (bottom) of the HAXTEL telescope concept. The gray bars give the normalized counts levels.

## 1.1 Focal plane detector requirements

The characteristics of the HAXTEL telescope configuration considered will require a focal plane detector able to fulfill the following requirements:

1. high stopping power: the detector has to be composed of a high-Z, high density material in order to provide significant stopping power at energies up to 600 keV and, therefore, a high detection efficiency over the entire operating range of the lenses;
2. spatial resolution: a spatial resolution of the order of ~1-2 mm in each direction is required;
3. energy resolution: the energy resolution is a critical scientific goal, a few keV at 500 keV;
4. small sensitive area: the area of the detector depends on the FOV. A trade off between low background and FOV should be implemented. A figure of 70 $cm^2$ as detection area is compatible with the reference configuration of the HAXTEL telescope.
5. sensitivity to the photon linear polarization [Lei et al., 1997]: this is a crucial requirement because several of the astrophysical emission processes of hard X–/soft gamma–ray radiation are expected to produce polarized photons. This capability is strictly related to the pixel size, which represents the geometrical spatial resolution of the detector.

## 2. CZT focal plane detectors

We have chosen CdTe/CZT as the starting material to design the focal plane detectors since it has emerged as a good material choice for the detection of hard X-rays and soft gamma-rays due to several characteristics: (a) the high detection efficiency [the high density of CZT (5.76 g/cm ) and high average atomic number (50) result in a high stopping power], (b) good spectroscopic performances, (c) good time response, all without the need for cryogenic cooling. Unfortunately the low mobility of the charge carriers (particularly the holes) and trapping/detrapping phenomena degrade the CdTe/CdZnTe detectors' response, depending on the distance between the charge formation position and the collecting electrodes. However the spectral properties can be improved by using either hardware or software techniques.

Because of their characteristics the CdTe/CZT semiconductors are boosting the field of hard X-ray astronomy. Currently there are two satellite that have onboard large CdTe/CZT detectors.

The International Gamma-Ray Astrophysics Laboratory (INTEGRAL) launched in 2002 carries the Imager on Board the Integral Satellite (IBIS) that uses a CdTe array with a sensitive area of 2600 cm$^2$. The operating energy range is 15 keV–1 MeV. After 2 years of flight operations, the 16384 CdTe crystals are very stable and only ~2.5% have been disabled due to their noisy behaviour. Moreover a gain degradation of ~2.6 % per year has been determined [Lebrun et al., 2005]. The observed background level allows a expected sensitivity of ~ 1milliCrab for a 10$^6$s observation and of ~ 1-2 10$^{-8}$ erg s$^{-1}$ cm$^{-2}$ in the 15 keV to 150 keV energy range ;

The Swift MIDEX spacecraft, launched in November 2004, with onboard the Burst Alert Telescope (BAT) equipped with a large area detector array to detect gamma-ray bursts (GRBs) and also to perform an all-sky hard X-ray survey with a limit sensitivity of ~2 mCrab. The detection area is of 5240 cm$^2$ and the detector plane is composed of 32768 pieces of CZT having the same dimensions of ISGRI pixels (4x4x2mm$^3$). The BAT operates in the 15-150 keV energy range with 6.2 keV resolution at 122 keV, a expected sensitivity of ~10$^{-8}$ erg s$^{-1}$ cm$^{-2}$ and a 1.4 sr (halfcoded) FOV [Barthelmy et al., 2005; Sato et al., 2005]. In twelve months of operation, BAT has detected and located onboard ~96 bursts.

## 2.1. CdTe/CZT Detectors: state of the art

Various efforts have been made to overcome the hole-trapping drawback in CdTe/CZT detectors, particularly relevant for thick crystals, which are needed for detecting hard X and gamma rays: these include the application of a grid structure to improve the energy resolution [Luke, 1994], the use of hemispheric or coaxial detectors and the compensation of the deficit of the pulse height based on the pulse rise time information [Lebrun et al., 1996; Verger et al., 2001].

An example of the spectroscopic response obtained with typical detecting materials is shown in figure 2: a) the detector tested is a Spectrometer Grade CZT single crystal (eV-Products, USA) of size 10×10×5 mm$^3$, with platinum contacts and grown by the High Pressure Bridgman technique. The 59.54 keV line of $^{241}$Am has a width (FWHM) of 4.2 keV and the energy resolution (FWHM) of the peak at 122 keV of $^{57}$Co is 4.8 keV; b) the detector has size 10×10×2 mm$^3$ (SMI-LAB, Ukraine). The energy resolution (FWHM) is 5.4 keV at 59.54 keV and 6.6 keV at 81 keV.

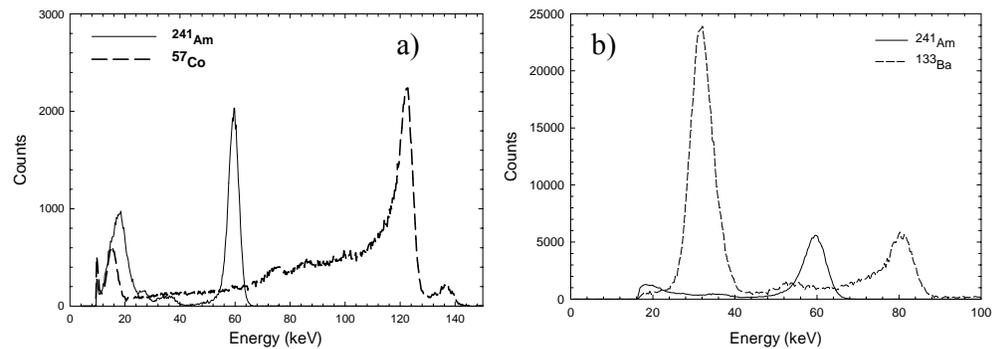

Fig. 2. a) Spectrum of $^{241}$Am and $^{57}$Co obtained with a CZT detector (Spectrometer Grade, 10 x 10 x 5 mm$^3$); b) Spectrum of $^{241}$Am and $^{133}$Ba recorded with a CZT detector (SMI-LAB, Ukraina, 10×10×2 mm$^3$).

## 2.1.1 Spectroscopic performances

Recently some improvement has been achieved in the spectroscopic performances utilizing indium as the anode electrode for p-type CdTe semiconductors [Tanaka et al., 2004]. In this way a high Schottky barrier formed on the In/p-CdTe interface allows the detector to operate as a diode and not as an ohmic device (e.g. as in the standard configuration with both Pt contacts). This leads to an improvement due to the very low leakage current achievable with the Schottky CdTe diode: at 400 V the leakage current of the In/CdTe/Pt detector with a thickness of 0.5 mm in the reversed biased condition is 2 orders of magnitude smaller than the leakage current of Pt/CdTe/Pt. As a consequence it is possible to apply a high electric field to ensure a complete charge collection. In terms of energy resolution, the In/CdTe/Pt detectors show excellent performances with respect to devices realized with Pt electrodes, as can be seen in fig. 3, where we presented some spectra acquired by us. However, at room temperature, they present time stability problems, since their original good spectroscopic performance degrades after hours. Another drawback is that the thickness of the crystal is currently limited to 1 mm.

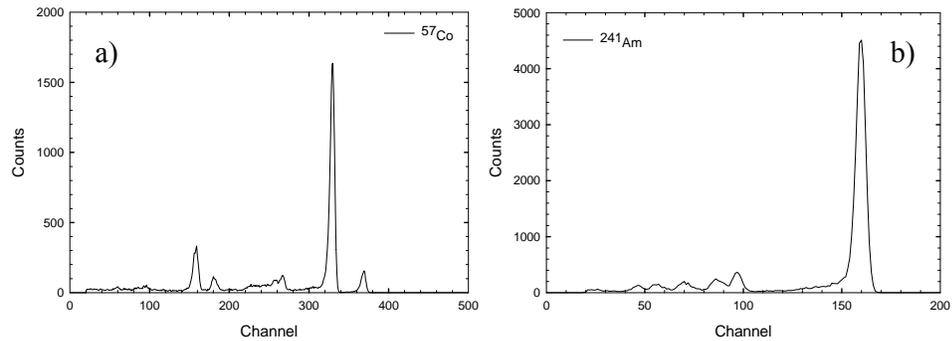

Fig. 3. Spectrum of $^{57}$Co (a) and $^{241}$Am (b) obtained with a Schottky CdTe detector (ACRORAD, Japan) of thickness 0.5 mm, biased at 400 V. The FWHM is 2% at 122 keV and 3.7% at 60 keV.

A direct technique to improve the spectroscopic performances is represented by detector cooling (see fig. 4a), while the method based on rise time discrimination has long been used to improve the spectral shape by using rise time information: all pulses with a rise time exceeding some threshold are rejected, in this way all pulses which would contribute to the trapping tail in the pulse height spectrum are rejected. This technique is effective in improving the resolution energy but the detector sensitivity, since many counts are discarded, is significantly reduced. We have tested the XR-100T-CZT (Amptek, USA), that is a high performance X-ray and gamma ray detection system. It is based on a planar CZT detector, mounted on a thermoelectric cooler inside a small hybrid package. The plot in fig. 4b shows the comparison between the spectra recorded with rise time discrimination turned off (FWTM=5.5 keV) and turned on (FWTM= 1.9 keV).

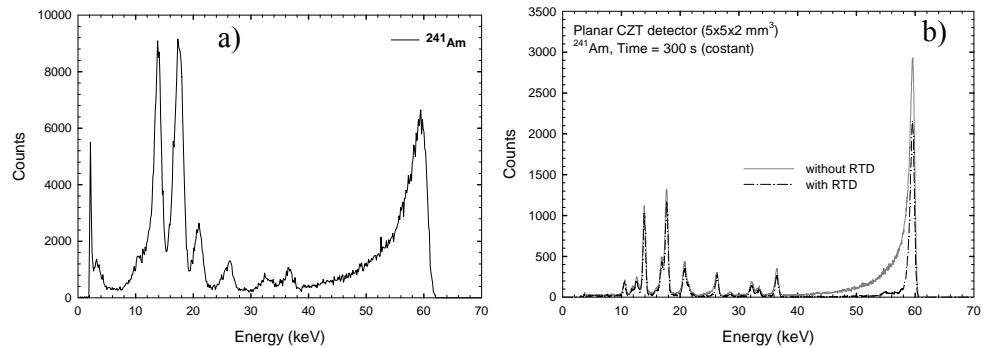

Fig. 4. a) $^{241}$Am spectrum acquired at -30 °C irradiating a CdTe detector (SMI-LAB, Ukraine) of size 5 x 5 x 2 mm$^3$. The energy resolution (FWHM) is 3.9 keV; b) Spectra of $^{241}$Am measured with an XR-100T-CZT (5x5x2 mm$^3$) with rise time discrimination turned off (FWHM= 0.85 keV) and turned on (FWHM= 0.76 keV). The spectra were accumulated for the same time (300 s).

The methods based on measuring both risetime and induced charge are termed biparametric methods. We are developing a biparametric method based on a hybrid hardware and software technique utilizing a double pulse shaping active filter (one [fast] compatible with the electron transit time, and the other [slow] to allow for complete charge collection) in order to analyze the detector signals and to obtain an indirect measurement of the interaction depth (direction z) between the electrodes of an incident photon in the detector material [Auricchio et al., 2004]. The pixel number identifies the interaction position in the x and y direction [Zhang et al., 2005].

Alternatively, a group at the Danish National Space Centre has recently developed a device that consist of an array of anodes on one side of a CZT (eV-Products) slab with a contiguous single cathode on the other side of the slab. In order to achieve a 2D imaging capability, the drift strip electrode geometry [Van Pamelen et al., 2000] was applied for the pixels. This arrangement, consisting of 16 drift detector cells, is shown in Fig. 5. Each drift detector comprise 2 drift electrodes and one anode readout electrode. The pixel pitch is 2 mm while the crystal size is 10 mm x 10 mm x 3 mm. The anode readout electrode radius is 0.2 mm. The inner and outer radii for the first surrounding drift electrode (G4) are 0.5 mm and 0.6 mm, respectively, while the rectangular common drift electrode (G3) is of 0.1 mm width. The first drift electrodes (G4) are interconnected externally by the z-bonding technique. The typical spectroscopic response of a pixel is shown in Fig. 6: the results are quite good, being about a factor two better than that achievable with monoelectrode detectors of equivalent area.

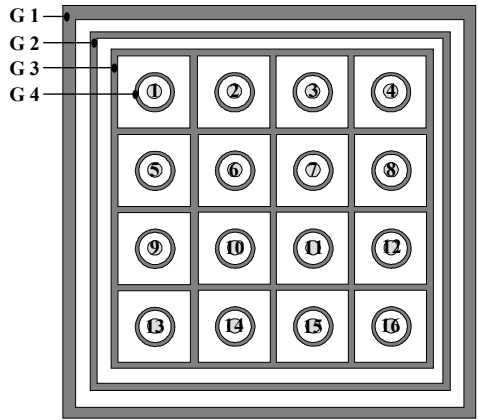

Fig. 5. Schematic design of the detector anode configuration (not to scale). The size of each pixel is 2×2×3 mm$^3$ [Kuvvetli and Budtz-Jørgensen, 2005].

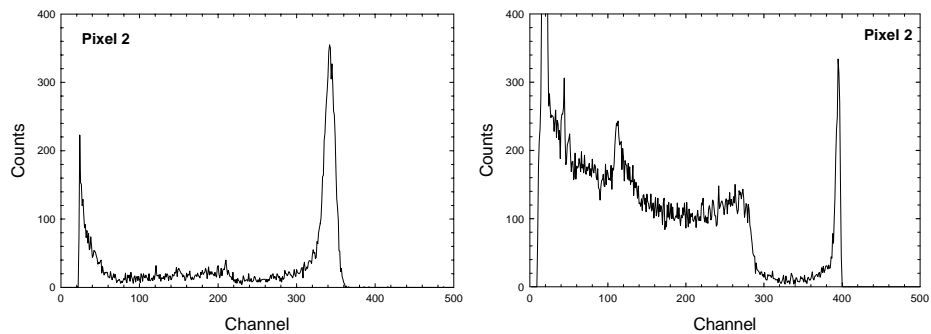

Fig. 6. $^{241}$Am spectrum (left): FWHM =2.7 keV; $^{137}$Cs spectrum (right): FWHM =7.8 keV.

*2.1.2 Response uniformity*

An important characteristic of a detector, in particular for focal plane implementation, is the response uniformity (in terms of gain, efficiency and energy resolution). In this aspect, one of the most critical roles is played by the quality and homogeneity of the CZT/CdTe crystals over large volumes. In Table I we have considered the performances of a large area CZT detector (IMARAD), 40×40 mm$^2$, 5 mm thick, composed by 16×16 pixels of 2.5×2.5 mm$^2$ area, and a 4×4 pixels CZT Multipixel (eV Products). The thickness of this detector is 5 mm, while the lateral dimensions are 10.6×10.6 mm$^2$ with a pixel size of 2×2 mm$^2$.

Table I: Response uniformity results from the CZT pixel detectors mentioned in the text.

|  | **IMARAD** (256 pixel) | **eV Multipixel** (16 pixel) |
|---|---|---|
| $\Delta E_{FWHM}/E$ (mean value) @ 122 keV | 4.8 %±0.64 (~10 %) | 2.82%±0.27 (~10 %) |
| Efficiency Dispersion from mean value (1 σ) @ 122 keV | ~4.5 % | ~ 8 % |

The non uniformity level shown by current devices can be easily corrected with an accurate calibration of each pixel, that involves the correction of the shift of photo-peak due to the charge transport properties and the readout chains gains equalization.

## 2.2. A Laue lens focal plane concept

An adequate detection efficiency can be obtained using devices based on CdZnTe with a thickness of more than 5-7 mm as shown in Fig. 7. It is, however, difficult to increase the thickness of the device while at the same time keeping good energy resolution throughout the energy range from 10 keV to a few hundred keV. This is because the effect of incomplete charge collection becomes more severe due to the low mobility and short lifetime of holes in CdZnTe. The inactive region in the detector volume degrades the energy resolution through a low energy tail in the spectral response.

In order to achieve high efficiency over the entire range, we propose a stacked CdTe/CZT pixel detector. Taking as reference the HAXTEL concept, a possible configuration of the focal plane detector for LEL is based on a CdTe/CZT pixellated layer, composed of 7 hexagonally shaped crystals (4 cm in diagonal) in order to minimize the detector surface required to cover the lens field of view. The pixels of each detector unit have the same geometry so as to provide the best coupling with the lens point spread function and to increase the symmetry for polarimetric studies. One mosaic layer of CdTe/CZT crystals could be the focal plane detector. The LEL focal plane (Fig. 8) can be considered as the basic layer with which to build the HEL focal plane stacked detector [Watanabe et al., 2005]: e.g. a three layer detector can be considered to have ~2 cm thickness and therefore to achieve a total detection efficiency above 60% up to 600 keV. Each layer can both operate as a separate position sensitive detector and polarimeter or work with other layers in order to enhance the total detection efficiency. The specifications of each single layer detector are reported in table II, while in Table III a tentative summary of the resources budget required both for LEL and HEL focal planes is given. As an alternative for the HEL instrument we can consider as the focal plane detector a single thick crystal used in the Planar Transverse Field configuration [Auricchio et al., 1999], where the applied electric field is orthogonal to the optical axis in order to obtain high detection efficiency, while maintaining small the charge collection distance so as to limit the trapping phenomena and to achieve a good energy resolution [Caroli et al., 2000].

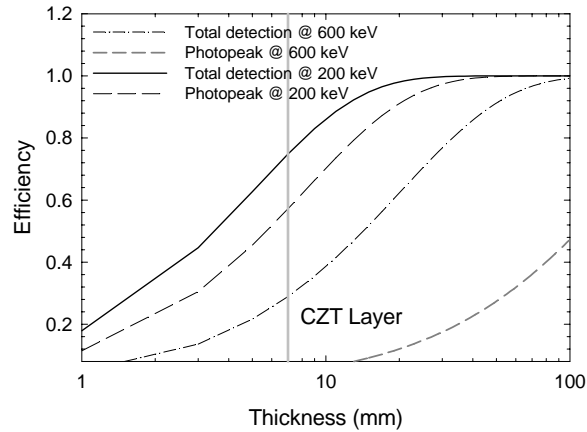

Fig. 7. CZT efficiency calculated versus detector thickness.

Table II. Single layer focal plane specifications.

| No. of units (crystals) | 7 |
|---|---|
| Detection area/unit | 10 cm$^2$ |
| Thickness | 7 mm |
| Pixel size | ~2mm ($\varnothing$) |
| Number of total channels | ~2700 |
| Total Detection Efficiency | 100% @ 60 keV<br>30% @ 600 keV |
| $\Delta E_{FWHM}/E$ | ~ 4.5 % @ 60 keV<br>~ 1 % @ 600 keV |
| Operating bandpass energy (keV) | 10 - 1000 keV |

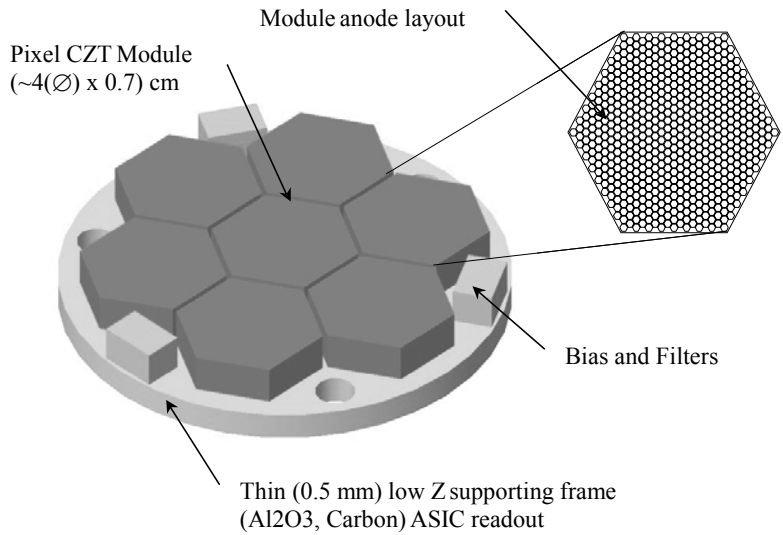

Fig. 8. Conceptual drawing of our layer module (used as LEL focal plane). For HEL three (or more) of this CZT layers could be stacked to achieve the required efficiency.

Table III. Resources budget for LEL and HEL Focal Plane.

| LEL focal plane | CZT Layer |
|---|---|
| Sensitive volume (CZT) | 50 cm$^3$ |
| Weight | 0.5 kg |
| Power | 3 W |
| Dimensions (Diameter×Height) | 12 cm × 1.5 cm |
| **HEL focal plane** | **3 CZT Stack** |
| Sensitive volume (CZT) | 150 cm$^3$ |
| Weight | 2 kg |
| Power | 10 W |
| Dimensions (Diameter × Height) | 12 cm × 5 cm |

# 3 Alternative focal plane detectors

CZT detectors satisfy the science requirements of a focal plane detector for LEL, promising excellent energy and spatial resolutions, and in addition the stacked detectors have high efficiency for HEL. Nevertheless it is convenient to consider alternative detector technologies, in relation to the cost and complexity associated with pixellated CZT detectors and particularly with their response uniformity.

Progress is being made in the development of new inorganic scintillators that, when coupled with wavelength-shifting fibers, photomultiplier tubes (PMTs) or photodiodes, could be successfully used to produce high performance hard-X ray detectors. Lanthanum bromide ($LaBr_3$) is a new inorganic scintillator material, that provides an improved energy resolution (R=3% FWHM at 662 keV) compared to CsI (R=7% FWHM at the same energy). When doped with cerium it is an attractive and promising candidate for high energy photon detection due to its high light output, high stopping efficiency, fast response, good linearity, and low cost. A valid alternative for focal plane detectors may be given by detection planes formed with LaBr3 scintillator coupled with a new generation of readout devices [McConnell et al., 2004].

The plots of fig. 9 compare the achievable efficiency considering configurations of focal plane detectors based on different materials and material combinations (reference thickness = 2 cm). As a reference these configurations are compared with a thick (8 cm) Ge detector such as that foreseen for a Laue Lens telescope proposed for Nuclear lines observation [Von Ballmoos et al., 2004]. Germanium detectors are semiconductor diodes having a p-i-n structure in which the intrinsic (i) region is sensitive to incident radiation. Advances in manufacturing techniques have allowed to grown extremely pure Ge crystals. Depletion depths of several centimeters can be achieved using this *High-Purity Germanium,* that also has the advantage to be stored at room temperature. HPGe detectors are operated at temperatures of around 77 K, in order to reduce noise from charge carriers, which can be thermally excited at room temperature across the relatively low band gap (0.67 eV); this induced noise destroys the energy resolution of the detector.

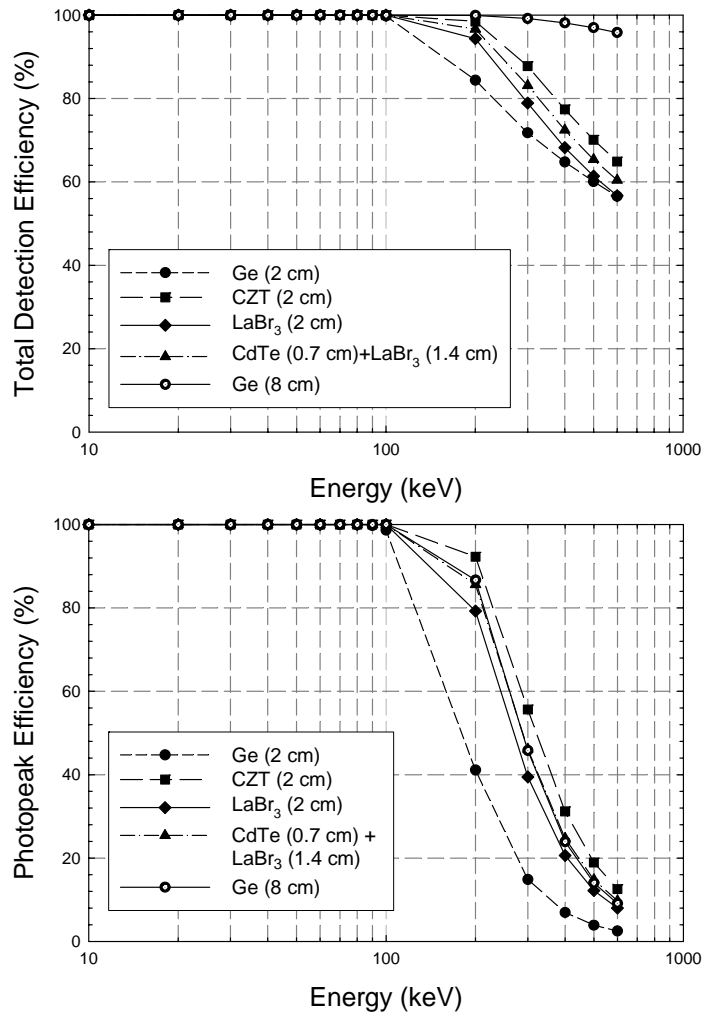

Fig. 9. Total detection efficiency and photopeak efficiency of the different configurations calculated as a function of the energy.

The characteristics of focal plane detectors based on different materials and/or material combinations are summarized in the following table:

Table IV. Main characteristics of different focal plane detectors.

| Detector type | Energy resolution FWHM (%) | | Spatial Resolution (mm) | Total Detection Efficiency (%) at 600 keV | Photopeak Efficiency (%) at 600 keV |
|---|---|---|---|---|---|
| | *122 keV* | *662 keV* | | | |
| Ge (2 cm) | ~0.6 | ~0.2 | 2 | 57 | 2.5 |
| Scintillators (LaBr$_3$-2 cm) | 6.8 | 2.7 | 1-2 | 57 | 8.0 |
| CZT stack (2 cm) | 3.5 | 1 | 1 | 65 | 12.6 |
| CZT + LaBr$_3$* (0.7 cm + 1.4 cm) | 3.5+6.8 | 1 + 2.7 | 1 + 1-2 | 60 | 9.7 |

*The total detection efficiency has been calculated in the last case as: $I=I_0 \exp(-(\mu_{CdTe} 0.7 + \mu_{LaBr3} 1.4))$

## 4 Conclusion

A detector suitable for space experiments should have high atomic number, high density and a band gap much larger than that of Ge to permit room temperature operation, without an expensive cryogenic cooling. The development of high performance CZT/CdTe pixel detectors offer a very good answer to the requirements of new high sensitivity telescopes for hard X-/soft γ-ray astronomy because CdTe combines relatively high atomic numbers (48 and 52) with a large bandgap energy (1.47 eV) and a density (~6 g/cm$^3$) higher than that of NaI and CsI. In particular this kind of detector would allow high detection efficiency over a wide operative energy range (e.g 60-600 keV) with small material volume, enabling a compact and effective shielding. This is important in order to reduce the locally produced background, which above 100 keV becomes the most important noise contribution. A further advantage is the quite low difficulty to produce highly segmented detectors that could also be very useful to improve background rejection (i.e. through some Compton kinematics event selection) and to obtain high polarimetric capabilities together with spectroscopic imaging and timing without losing detection efficiency.

In order to achieve high efficiency over the entire range 60-600 keV, we propose a stack of CZT pixel devices suitable as focal plane detector for Laue lens telescopes as the HAXTEL concepts, but before defining optimum instrument configuration we have planned a study of the instrumental background in different orbital radiation environment.

### Acknowledgments

The authors wish to acknowledgment Alessandro Traci and Damiano Pellicciotta for their help on the preparation of focal plane drawings and laboratory measurements..